\documentclass[aps,prc,twocolumn,twosided]{revtex4}
\usepackage{epsfig,amsmath}

\begin{document}

\title{Large mass dileptons from the passage of jets through quark gluon plasma}
\author{Dinesh K.~Srivastava}
\altaffiliation{on leave from: Variable Energy Cyclotron Centre, 
             1/AF Bidhan Nagar, Kolkata 700 064, India}            
\affiliation{Physics Department, McGill University,
             3600 University Street, Montreal, H3A 2T8, Canada} 
\author{Charles Gale}
\affiliation{Physics Department, McGill University,
             3600 University Street, Montreal, H3A 2T8, Canada} 
\author{Rainer J.~Fries}
\affiliation{Department of Physics, Duke University, 
             Durham, NC 27708-0305, USA}
\date{\today}
\begin{abstract}
We calculate the emission of large mass dileptons originating from the
annihilation of quark jets passing through quark gluon plasma.
Considering central collisions of heavy nuclei at SPS, RHIC
and LHC energies, we find that the yield due to the jet-plasma
interaction gets progressively larger as the
collision energy increases. We find it to be
 negligible at SPS energies,
of the order of the Drell-Yan contribution and much larger than
the normal thermal yield at RHIC energies and
up to a factor of ten larger than the Drell-Yan contribution at
LHC energies. An observation of this new dilepton source would confirm the
occurrence of jet-plasma interactions
and of conditions suitable for jet-quenching to take place.
\end{abstract}
\maketitle

\section{Introduction}

Collisions of relativistic heavy ions are performed  with a hope to create 
a plasma of quarks and gluons (QGP) and to study  the
deconfined state of strongly
interacting matter. Theoretical investigations of the formation, evolution
and signatures of the QGP have played an important role in planning the
experiments for these studies. One such investigation concerns 
the radiation of dileptons which 
are considered as penetrating probes of such matter due to their
very long mean free path. Several sources of large mass dileptons have been
identified and studied. These include the radiation of Drell-Yan pairs
from the annihilation of primary partons in the nuclei~\cite{dy}, 
thermal dileptons from the annihilation of
quarks and antiquarks in the plasma~\cite{fs},
and from hadronic reactions in the hot hadronic gas~\cite{gale} following the 
hadronization of the plasma.
Dileptons from the correlated decay of charm and bottom quarks have also
been considered~\cite{ramona}.
Dileptons from the decay of hadrons like
$\rho$ and $\phi$ will provide information about the possible
medium modification of hadrons.
The pre-equilibrium production of dileptons has also been investigated by
several authors in various 
approximations~\cite{el:90, gk:93, kae:94, str:94, SriMusMul:97}.

Experimentally the low mass dilepton excess observed in the
nucleus-nucleus collisions at SPS energies~\cite{pbpb}
has lead to suggestions that interactions among hadrons, and perhaps
their properties as well,
are altered in hot and dense nuclear matter~\cite{rapp1}. The
interpretation of the excess observed in 
intermediate mass dileptons~\cite{na50} has been done in terms of a  thermal hadronic environment
with some quark-gluon plasma component~\cite{SriSiG:96,rapp2,KGS:02}.
The measurement of dileptons has already made a considerable progress
at RHIC experiments~\cite{rh_dil}.

In the present work, we discuss the emission of large mass dileptons
induced by the passage of high energy quark-jets through the QGP. 
The general formulation for their emission rate along with
 with a discussion of the evolution of the plasma, including the possible
chemical equilibration, and the spectrum of jets and Drell-Yan
dileptons, is given in Sect.\ II.
The numerical results and a discussion are given in Sect.\ 
III followed by a summary.

\section{Formulation}

Let us consider  a collision of two heavy nuclei at relativistic energies.
 A large number of minijets produced in the collision
and the subsequent rescattering of the partons will lead to the formation of
a thermalized plasma at some initial temperature
$T_0$ at time $\tau_0$. This plasma may or may not be in chemical equilibrium.

The high energy  partons, which are produced at times 
$\tau \sim {1/p_T} \ll \tau_0$, will traverse through this plasma, losing 
energy through collisions and radiation of gluons, a phenomenon widely 
known as jet-quenching and considered to be an important
probe of the QGP \cite{GyulWang:94}. 

The quark jets passing through the plasma will produce
large mass dileptons by annihilation with  the thermal antiquarks.
They will also produce high energy photons~\cite{fms}.
If these additional dileptons and photons are detected, they will provide 
a confirmation for the formation of the plasma and the occurrence of 
jet-plasma interactions and prepare the ground for a deeper understanding of 
the jet-quenching phenomenon and the evolution of
the colliding system. We shall see that the production is quite sensitive
to the temperatures at early times and controlled by the
initial distribution of high energy partons.

In order to proceed we assume that the phase space distribution of the quarks
in the medium can be decomposed as
\begin{equation}
f({\bf{p}}) =f_{\text{th}}({\bf{p}})+ f_{\text{jet}}({\bf{p}})
\end{equation}
where we have a thermal component that is characterized by a temperature $T$,
\begin{equation}
f_{\text{th}}({\bf{p}})=\exp(-E/T).
\end{equation}
and a jet component which is assumed to be
limited to partons having a $p_T \gg$ 1 GeV. This separation is kinematically reasonable, as the 
jet spectra fall off as a power law and can thus easily be differentiated from thermal components. 
We shall, somewhat arbitrarily, limit them to having $p_T \ge $ 4 GeV.
In addition we give some results when only jets with $p_T \ge $ 6 GeV are 
considered.

 The phase-space distribution for the quark-jets is given by the
 perturbative QCD result for the jet yield as
\cite{LinGyu:95}
\begin{multline}
  f_{\rm jet}({\bf p})=\frac{1}{g_q}\frac{(2\pi)^3}
  {\pi R_\perp^2 \, \tau \, p_T \cosh y} \,
  \frac{dN_{\rm jet}}{d^2p_T dy} \, \\[0.5em]   \times
  \delta (\eta-y) \, \Theta(\tau-\tau_i) \, \Theta (\tau_{\rm max} - \tau) 
\, \Theta(R_\perp - r)  \>.
\label{fjet}
\end{multline}
where $g_q=2\times 3$ is the spin and colour degeneracy of the
quarks, $R_\perp$ is the transverse dimension of the system, 
$\tau_i \sim 1/p_T$ is the formation time for the jet and $\eta$ is the 
space time rapidity.
We shall take $\tau_{\rm max}$ as the smaller of the lifetime of 
the QGP ($\tau_f$) and
the time taken by the jet produced at position $\bf r$ with  a velocity 
$\bf v $ in the transverse direction to reach the surface of the plasma. 
The distance covered by the parton during this passage is given by
\begin{equation}
  d=-r \cos \phi + \sqrt{R_\perp^2 -r^2 \sin^2 \phi}
\end{equation}
where $\cos \phi= \widehat{\bf v} \cdot \widehat{\bf r}$. We assume here that
the parton is massless and travels with the speed of light.
The boost invariant Bjorken correlation \cite{Bj:83} between the rapidity $y$ and 
$\eta$ is assumed.

These phase space distributions can be folded with the cross section
of the annihilation of the quarks to obtain the rate of emission of the
dileptons. The rates are then convoluted with the space time history
of the evolution of the plasma, using a standard procedure. 

\subsection{Rate of emission of dileptons}

The cross section for quark-antiquark annihilation into dileptons
($a b \rightarrow \ell^+ \ell^-$) is given by
\begin{equation}
  \sigma(M^2)= \frac{4\pi}{3}\,\frac{\alpha^2}{M^2}\,N_c (2s+1)^2 \, 
  \sum_f e_f^2 
\end{equation}
where the sum runs over the flavour of quarks, $N_c=3$, and $s$ and $e_f$
stand for the spin and the charge of the quark.

Using kinetic theory the reaction rate for the above process can 
be written as
\begin{equation}
R=\int \frac{d^3p_a}{(2\pi)^3} f_a({\bf{p}}_a) \int \frac{d^3p_b}{2\pi^3}
  f_b({\bf{p}}_b) \sigma(M^2) v_{\text{rel}} \>,
\end{equation}
where $f_i$ stands for the phase-space distribution of the quark or the
antiquark, ${\bf{p}}_a$ and ${\bf{p}}_b$ are their momenta respectively 
and the relative velocity is
\begin{equation}
v_{\text{rel}}=\frac{E_a E_b -{\bf{p}}_a {\bf{p}}_b}{E_a E_b} \>.
\end{equation}

After some algebra~\cite{kms:92} this can be rewritten as
\begin{widetext}
\begin{equation}
\frac{dR}{dM^2}=\frac{M^6}{2}\frac{\sigma(M^2)}{(2\pi)^6} \int
x_a \, dx_a \, d\phi_a \, x_b \, dx_b \, d\phi_b \, dy_a \, dy_b
 f_a\, f_b \>
\delta\left[M^2-2M^2 x_a x_b \cosh (y_a-y_b)+2 M^2 x_a x_b \cos \phi_b\right]
\end{equation}
\end{widetext}
where $x_a=p_T^a/M$, $x_b=p_T^b/M$ and $y_a$ and $y_b$ are the rapidities.
The integrations over the azimuthal angles yield
\begin{multline}
\frac{dR}{dM^2}=\frac{M^4\sigma(M^2)}{(2\pi)^5} \int
x_a \, dx_a \, x_b \, dx_b \, dy_a \, dy_b\, 
 f_a \,f_b \, \\ \left [ 4 x_a^2 x_b^2 -
\left\{2 x_a x_b \cosh (y_a-y_b)-1\right\}^2
\right ]^{-1/2}
\label{gen}
\end{multline}
such that,
\begin{eqnarray}
 -1 &  \le & \, \frac{2x_a x_b \, \cosh (y_a-y_b) -1}{2 x_a x_b} \, \le \, 1
\> , \nonumber\\
 0 & \le & \, x_{a,b}\, \le \,  \infty \> , \\
-\infty & \le & \, y_{a,b} \, \le \, \infty \> . \nonumber
\end{eqnarray}
When $f_a$ and $f_b$ are given by a thermal distribution
\begin{equation}
f_{\text{th}}({\bf{p}})=\exp(-E/T)=\exp(-p_T \cosh y/T)
\label{fther}
\end{equation}
the above integral can be performed \cite{KKMcLM:86} to obtain  
\begin{equation}
\frac{dR}{dM^2}=\frac{\sigma(M^2)}{2(2\pi)^4} M^3 T K_1(M/T) \>.
\label{ther}
\end{equation}

The expression in Eq.~(\ref{gen}) can be used for phase space distributions
of arbitrary form.
When the distributions $f_a$ and $f_b$ are taken as thermal,
we shall call the yield as thermal and when one of them is thermal and
the other one is from a jet, we call the yield as due to the 
passage of jets through the plasma in an obvious nomenclature.
Noting that the jet-distribution (Eq.~\ref{fjet}) depends on the
quark flavour, we take charge-weighted average of the
distributions for $u$, $d$, and $s$ quark-jets and for the corresponding
anti-quark jets, while using $f_{\rm jet}$ in Eq.~(\ref{gen}):
\begin{equation}
f_{jet}^{q(\overline{q})}({\bf p})=
 \sum_{f} e_f^2 f_{\rm jet}^f ({\bf p})/
           \sum_{f} e_f^2 ~,
\end{equation}
where the sum runs over the flavours of quarks (anti-quarks). The thermal
distributions are assumed to be flavour-independent, so that
the relation (Eq.~\ref{fther}) is recovered for them.

\subsection{Evolution of the plasma}

The radiation of dileptons from an expanding and cooling plasma
 including the contributions from a mixed phase and the hadronic
phase with transverse expansion of the plasma has been
studied in great detail (see e.g.\ \cite{KGS:02}).
We are interested in the emission of dileptons having large masses
and thus it is sufficient to consider emissions from the QGP phase.  
We also ignore  the transverse expansion of the
plasma~\cite{SriMusMul:97}. 

\begin{table}[t]
\begin{tabular}{||c||c|c|c|c|c||}
  \hline\hline
  & $\tau_0$ & $T_0$ & $\lambda_g^{(i)}$ &$ \lambda_q^{(i)}$&
 $\epsilon_i$\\  
 & (fm/$c$) & (GeV) & - & - & (GeV/fm$^3$)\\
\hline\hline    \multicolumn{6}{||c||}{Bjorken formula}  \\ \hline
SPS & 0.20 &0.345 & 1.0 & 1.0 &26.10 \\
& 0.50 &0.254 & 1.0& 1.0 &7.94 \\
\hline
RHIC &0.15 &0.447 & 1.0 & 1.0 &72.38 \\
     &0.50 &0.297 & 1.0 & 1.0 &14.43\\
\hline
LHC &0.073 &0.897 & 1.0 & 1.0 &1172 \\
     &0.50 &0.473 & 1.0 & 1.0 &90.92\\
\hline   \multicolumn{6}{||c||}{Self-screened parton cascade}  \\ \hline
RHIC  & 0.25 & 0.67 &0.34 & 0.064 & 61.4 \\ \hline 
LHC & 0.25  & 1.02 & 0.43 & 0.082 &425 \\
   \hline\hline
\end{tabular}
  \caption{Initial conditions for the hydrodynamical expansion estimated 
  from multiplicity densities using the Bjorken formula and from a 
  self-screened parton cascade model \cite{sspc}.}
  \label{tab:sspc}
\end{table}


As a first step, we assume that a thermally and chemically equilibrated plasma
is produced in the collision at time $\tau_0$. Assuming furthermore 
an isentropic expansion \cite{Bj:83,HwaKaj:85} one can write that
\begin{equation}
\frac{2\pi^4}{45\zeta(3)}\frac{1}{A_\perp} \frac{dN}{dy}=4 a T_0^3 \tau_0
\end{equation}
where $dN/dy$ is the particle rapidity density for the collision
and $a=42.25 \pi^2/90$ for a plasma of massless $u$, $d$ and $s$ quarks and 
gluons. We have taken the number of quark flavours as $\approx 2.5$ 
to account for the mass of $s$ quarks while evaluating $a$.
The transverse area of the system is taken as $A_\perp=\pi R_\perp^2$ where
$R_\perp=1.2 A^{1/3}$ is the transverse radius of the system for a
head-on collision. We assume a rapid thermalization limited by $\tau_0\sim
1/3\,T_0$ \cite{kms:92}.

We take the particle rapidity density as 750 for central collisions of
lead nuclei at SPS energies.
At RHIC energies we estimate the particle rapidity density as
 $\approx$ 1260, based on the  measured 
charged particle pseudo-rapidity density
for central collisions of gold nuclei at $\sqrt{s_{NN}}=$ 200 GeV. For 
central collisions of lead nuclei at LHC energies we use $dN/dy=5625$, 
as suggested in \cite{kms:92}.

The initial temperatures and times from these estimates are given in 
Table~\ref{tab:sspc} under the heading Bjorken formula. We also use an 
alternative estimate where the formation time is taken as 0.50 fm/$c$,
 so that the corresponding initial temperatures are considerably
reduced. We shall see that this drastically alters the yield of
dileptons from interactions within the plasma, while the 
yield due to the jet-plasma interaction  only reduces 
by a factor of about 2.

\begin{table*}[ht]
\begin{tabular}{||c||c|c|c|c|c||}
  \hline\hline
   &  & 
  \begin{minipage}{3cm}\begin{center} {$C \quad [1/{\rm GeV}^2]$}
  \end{center}
  \end{minipage} & 
  \begin{minipage}{3cm}\begin{center} {$B \quad [{\rm GeV}]$ }
  \end{center}\end{minipage} & 
  \begin{minipage}{3cm}\begin{center}{$\beta$}\end{center}
  \end{minipage} &   
  \begin{minipage}{3cm}\begin{center}{$\delta$}\end{center}
  \end{minipage}   \\ \hline\hline
  & u & 5.770$\times 10^5$ & 0.2657 & 6.638 & 7.682 \\ \cline{2-6}
  & d & 4.508$\times 10^5$ & 0.2904 & 6.703 & 7.631 \\ \cline{2-6}
  SPS & s & 2.760$\times 10^5$ & 0.3276 & 6.846 & 10.696 \\ \cline{2-6}
  & $\overline{\rm u}$ & 1.147$\times 10^5$ & 0.2826 & 6.781 & 10.949 
  \\ \cline{2-6}
  & $\overline{\rm d}$ & 6.028$\times 10^5$ & 0.1809 & 6.402 & 10.942 
  \\ \cline{2-6}
  & g & 3.180$\times 10^5$ & 0.5344 & 8.071 & 8.964 
  \\ \hline \hline
  & u & 9.113$\times 10^2$ & 1.459 & 7.679 & -- \\ \cline{2-6}
  & d & 9.596$\times 10^2$ & 1.467 & 7.662 & -- \\ \cline{2-6}
  RHIC & s & 1.038$\times 10^2$ & 1.868 & 8.642 & -- \\ \cline{2-6}
  & $\overline{\rm u}$ & 2.031$\times 10^2$ & 1.767 & 8.546 & -- 
  \\ \cline{2-6}
  & $\overline{\rm d}$ & 2.013$\times 10^2$ & 1.759 & 8.566 & -- 
  \\ \cline{2-6}
  & g & 4.455$\times 10^3$ & 1.7694 & 8.610 & -- \\ \hline\hline
  & \hspace{1em}u\hspace{1em} & 2.209$\times 10^4$ & 0.5635 & 5.240 & --
  \\ \cline{2-6}
  & d & 2.493$\times 10^4$ & 0.5522 & 5.223 & --
  \\ \cline{2-6}
  \hspace{1em}LHC\hspace{1em} & s & 1.662$\times 10^3$ & 0.9064 & 5.548 & --
  \\ \cline{2-6}
  & $\overline{\rm u}$ & 4.581$\times 10^3$ & 0.7248 & 5.437 & -- 
  \\ \cline{2-6}
  & $\overline{\rm d}$ & 4.317$\times 10^3$ & 0.7343 & 5.448 & -- 
  \\ \cline{2-6}
  & g & 1.229$\times 10^5$ & 0.7717 & 5.600 & -- \\ \hline\hline  
\end{tabular}
  \caption{Parameters for the  minijet distribution 
  $dN/d^2 p_T \, dy$ given in Eq.~(\ref{eq:para}) at $y=0$ for Pb+Pb at
  $\sqrt{s_{\rm NN}}=17.4$ GeV (SPS), Au+Au at 
  $\sqrt{s_{\rm NN}}=200$ GeV (RHIC) and for Pb+Pb at $\sqrt{s_{\rm NN}}=5.5$
  TeV (LHC). For all parameterizations $K=2.5$. 
  CTEQ5L plus EKS98 nuclear parton distributions are
  used. The range of validity is $p_T=$
  2 -- 20 GeV for RHIC and LHC and 2 -- 7 GeV for SPS.}
  \label{tab:minijets}
\end{table*}

We further approximate the nuclei as being spherical in shape 
with a uniform density and assume that the energy density at
the transverse position $r$ is proportional to the number of
collisions so that
\begin{equation}
\epsilon(\tau_0,r) \propto  4 \rho_0^2 \left[ R_\perp^2-r^2 \right] 
\end{equation}
where $\rho_0$ is the density of the nucleons. The normalization is then
determined by the condition that
\begin{equation}
A_\perp \epsilon_0=\int 2 \,\pi \, r \, \epsilon (r) \, dr
\end{equation}
where $\epsilon_0$ is the average energy density decided by the
temperature $T_0$. We note that this results in a transverse profile 
for the initial temperature as
\begin{equation}
 T(r)=T_0\left[ 2\left( 1-\frac{r^2}{R_\perp^2}\right) \right]^{1/4}.
\end{equation}
We  account for the transverse profile of the jet production as
well by introducing a factor $2(1-r^2/R_\perp^2)$  in expression (\ref{fjet})
while performing the space-time integration
\begin{equation}
d^4x=\tau d\tau \, r dr \, d\eta \, d\phi ~.
\end{equation}
The limits of the $\tau$-integration are $[\tau_0,\tau_f]$ for the 
purely thermal contribution. While estimating the yield of dileptons 
due to the passage of jets through the plasma the upper limit has to be 
taken as the 
minimum of $d/c$ and $\tau_f$. Assuming a Bjorken cooling
(for the chemically equilibrated plasma), so that
$T^3 \tau$ is constant,  we estimate $\tau_f$ to be
\begin{equation}
\tau_f=\left[\frac{T_0}{T_c}\right]^3 \tau_0 \> ,
\end{equation}
so that it is about 2 fm/$c$ at SPS energies, about 3.3 fm/$c$ at RHIC
energies, and about 13 fm/$c$ at LHC energies, when we take $T_c$ as 160 MeV,
and $T_0$ and $\tau_0$ from Table~\ref{tab:sspc}.
Thus we note that the 
average value for $d$
\begin{widetext}
\begin{eqnarray}
\langle d \rangle &=&\frac{3}{2\pi R_\perp}
\int_0^{2\pi}\, d\phi \, \int_0^{R_\perp}\, dr \> d \, \left(1-r^2/R_\perp^2
\right)\,   \nonumber \\[1.5em]
&=& \frac{ R_\perp}{2} \left[3\,{ }_{3\!}F_2
\left(-\frac{1}{2},\, \frac{1}{2},\, \frac{1}{2};
1,\, \frac{3}{2}; \,1\right)  
 - { }_{3\!}F_2\left(-\frac{1}{2},\,\frac{1}{2},\,\frac{1}{2};
1,\, \frac{5}{2}; \,1\right)\right] 
\> \approx \> 0.94 \, R_\perp
\end{eqnarray}
\end{widetext}
is smaller than the lifetime of the plasma only at LHC energies.
In the above, the ${}_p\!F_{q}$'s  are generalized hypergeometric functions.
Of course the transverse expansion of the plasma, which should be 
seen at LHC will reduce this lifetime \cite{first,SriMusMul:97} to some
extent.
Thus, on an average, the jet covers only a short distance in the hot
plasma, before the medium cools down to $T_c$ \cite{GKP:02}, except at LHC,
where the jets get to traverse the entire length of the system, while it is
still hot and deconfined. We shall see that this leads to
a large yield of dileptons from the jet-plasma interaction at LHC.

\subsection{Chemically equilibrating plasma}

Several model calculations have suggested (see e.g. Ref.~\cite{sspc})
 that while the plasma
to be created in relativistic heavy ion collisions may attain 
kinetic equilibrium fairly quickly, the evolution to the 
chemical equilibrium \cite{biro} proceeds only slowly through
reactions of the type $gg \leftrightarrow q\overline{q}$ and gluon 
multiplication and fusion $gg \leftrightarrow ggg$. Defining the
quark and gluon fugacities through
\begin{equation}
n_g=\lambda_g \tilde{n}_g \>, \> n_q= \lambda_q \tilde{n}_q \>,
\end{equation}
where $n_i$ are the actual densities and $\tilde{n}_i$ are the equilibrium 
densities for the parton species $i$, we can write
\begin{eqnarray}
\tilde{n}_g&=&\frac{16}{\pi^2}\zeta(3) T^3 =a_1 T^3 \>, \nonumber\\
\tilde{n}_q & = & \frac{9}{2 \pi^2} \zeta(3) N_f T^3 =b_1 T^3 \>.
\end{eqnarray}
The energy density and the pressure of the plasma are then given by,
\begin{equation}
\epsilon=3 P = \left[ a_2 \lambda_g +b_2 (\lambda_q +\lambda_{\overline{q}})
        \right ]T^4 \>,
\end{equation}
where $a_2=8\pi^2/15$, $b_2= 7 \pi^2 N_f/40$ and $N_f \approx 2.5 $ is the
number of dynamical quark flavours. We further assume that
 $\lambda_q=\lambda_{\overline{q}}$ . Assuming a boost invariant 
longitudinal expansion, the relevant master equations can be solved
to give 
\begin{eqnarray}
\frac{1}{\lambda_g}\frac{d\lambda_g}{d\tau} +\frac{3}{T}\frac{dT}{d\tau}
+\frac{1}{\tau}& = &R_3 (1 -\lambda_g) - 2 R_2 
\left ( 1 -\frac{\lambda_q \lambda_
{\overline{q}}}{\lambda_g^2}\right) , \nonumber\\[1em]
\frac{1}{\lambda_q}\frac{d\lambda_q}{d\tau}+\frac{3}{T}\frac{dT}{d\tau} + 
\frac{1}
{\tau}& = &R_2 \frac{a_1}{b_1} \left (\frac{\lambda_g}{\lambda_q} -
\frac{\lambda_{\overline{q}}}{\lambda_g}\right) ,
\end{eqnarray}
where the rate constants
\begin{eqnarray}
  R_2 &\approx& 0.24 N_f \alpha_s^2 \lambda_g T \ln ( 1.65/\alpha_s \lambda_g) 
\>, \\
  R_3 &=& 1.24 \alpha_s^2 T (2\lambda_g-\lambda_g^2)^{1/2} \>,
\end{eqnarray}
are taken from Ref.~\cite{biro} and include the effects of colour Debye
screening and Landau-Pomeranchuk-Migdal suppression of the
induced gluonic radiations. The initial conditions necessary for a
numerical solution of the above equation are taken from the
self screened parton cascade model~\cite{sspc} and given in 
Table~\ref{tab:sspc},
for the sake of completeness.

\subsection{Jets and Drell-Yan}

We estimate the jet production and Drell-Yan production of
dileptons in lowest order pQCD.
The jet cross section, normalized to one nucleon pair, 
for the production of partons (quarks, antiquarks or gluons) in primary hard 
interactions between 
partons $a$ and $b$ from different nuclei $A$ and $B$ 
is given by \cite{Owens:87}
\begin{multline}
  \frac{d\sigma^{\rm jet}}{d^2 p_T \, d y} = \sum_{a+b\rightarrow 
  {\rm jet}} 
  \int_{x_a^{\rm min}}^1 d x_a
  f_a^A(x_a)f_b^B(x_b) \\  \times
  \frac{x_a x_b \sqrt{s_{\rm NN}}}{x_a \sqrt{s_{\rm NN}} 
  - p_T e^y} \frac{1}{\pi}
  \frac{d\sigma_{a+b\rightarrow {\rm jet}}}{dt}.
  \label{eq:minijet}
\end{multline}
Here $f_a^A$ and $f_b^B$ are parton distributions for the colliding nuclei, 
so that
\begin{multline}
f_a^A(x,Q^2) =R_a^A(x,Q^2) \\[1.0em]  \times
 \left[ \frac{Z}{A} f^p_a (x,Q^2)+ \frac{N}{A} f^n_a (x,Q^2)\right], 
\end{multline}
where $R_a^A$ is the nuclear modification of the structure function, $Z$ is
the number of protons, $N$ is the number of neutrons and $Q^2=p_T^2$.
The sum runs over all
parton processes where a parton, forming a jet, can be produced from $a$ and 
$b$ with differential cross section $d\sigma /dt$.
$\sqrt{s_{\rm NN}}$ is the center of mass energy of the colliding nuclei 
per nucleon pair, so that $s=x_a x_b s_{\rm NN}$. Furthermore we have
\begin{eqnarray}
  x_b &=& \frac{x_a p_T e^{-y}}{x_a \sqrt{s_{\rm NN}} - p_T e^{y}} 
  \>, \\[1.0em]
  x_a^{\rm min} &=& \frac{p_T e^{y}}{\sqrt{s_{\rm NN}}-p_T e^{-y}} 
  \>.
\end{eqnarray}
\begin{figure}[tb]
  \begin{center}
  \epsfig{file=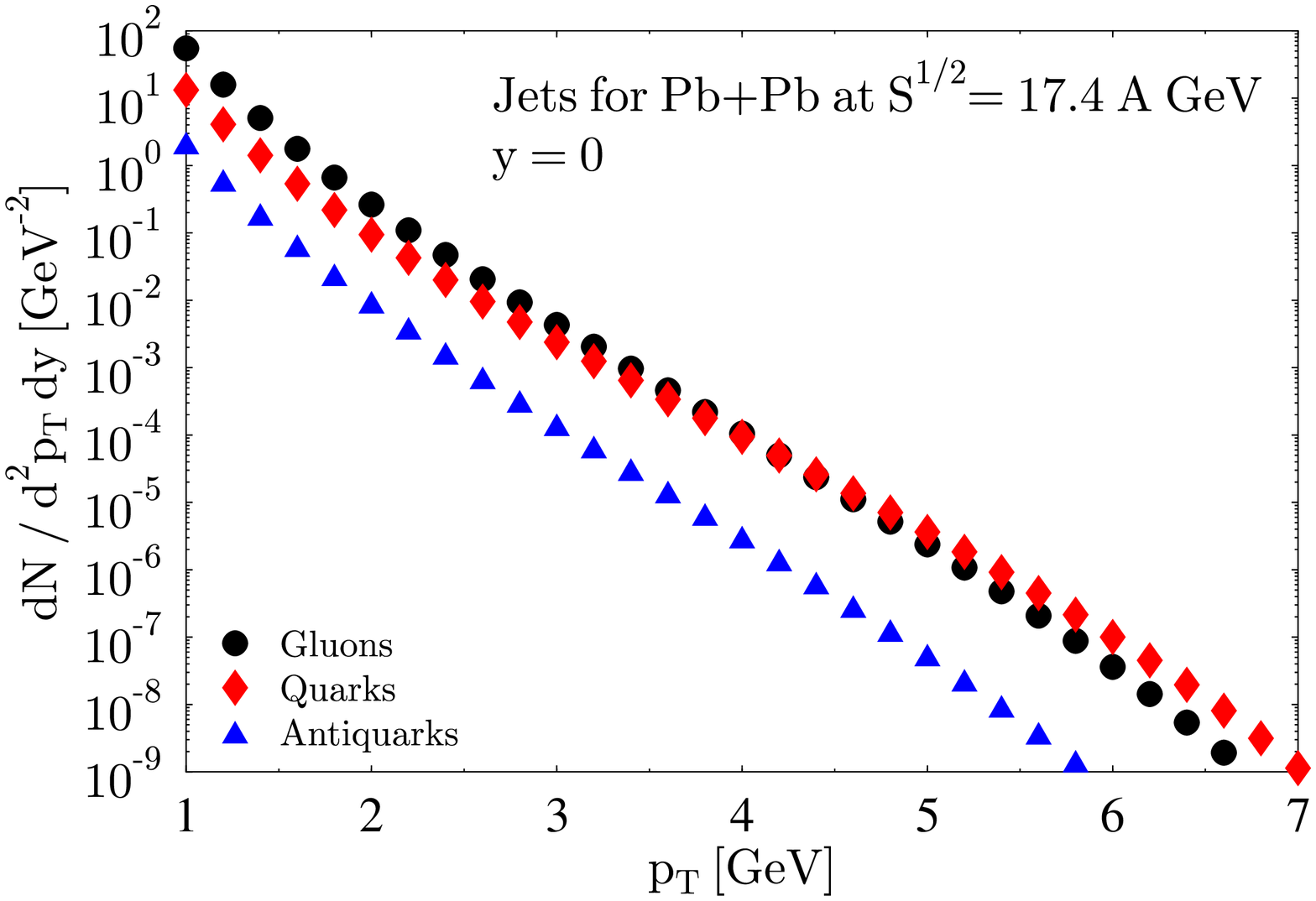,width=8.6cm} 
  \epsfig{file=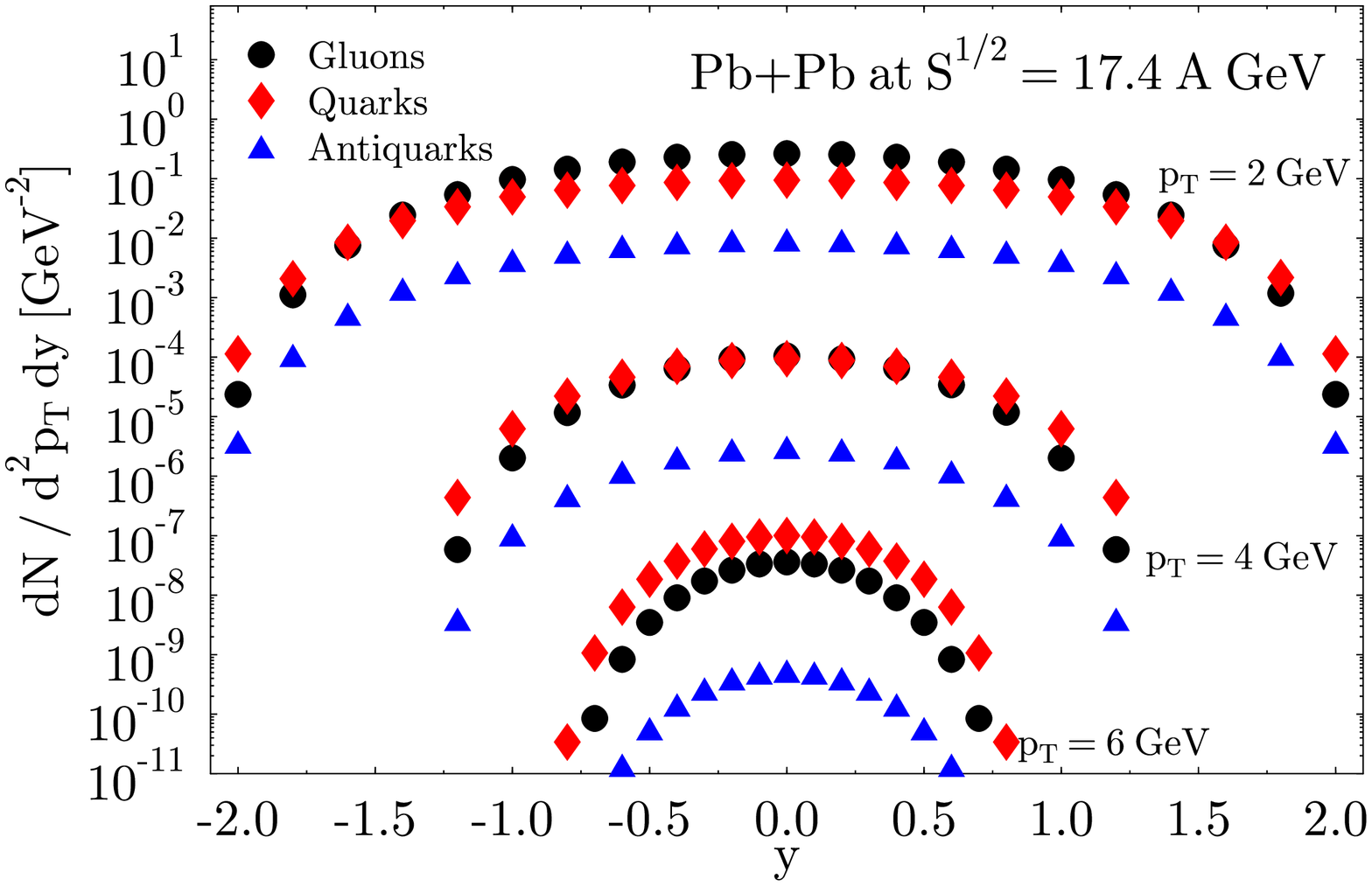,width=8.6cm} 
  \caption{Upper part: transverse momentum distribution of gluons, quarks and 
   antiquarks at $y=0$ for central Pb+Pb collisions at $\sqrt{s_{NN}}=17.4$ 
   GeV at SPS. Points are calculated, the lines are drawn as a
   guide to the eye. 
    Lower part: the rapidity distribution for
   gluons, quarks and antiquarks at SPS for $p_T=2$, 4 and 6 GeV 
   (from top to bottom). In this, and in the following two figures,
we have plotted the average of the three lightest flavours, $q=(u+d+s)/3$,
and $\overline{q}=(\overline{u}+\overline{d}+\overline{s})/3$, to
avoid overcrowding.}
  \label{fig:jetsps}
  \end{center}
\end{figure}
We have used the CTEQ5L 
parton distributions \cite{cteq5} and EKS98 nuclear modifications \cite{EKS98}.
We further use a $K$-factor of 2.5 in the jet production to account
for higher order corrections (see later).
The results of our calculations can be found in Figs.~\ref{fig:jetsps},
~\ref{fig:jetrhic} and
\ref{fig:jetlhc}. We also give a fairly accurate parameterization
for the results at RHIC and LHC energies;
\begin{eqnarray}
  \frac{d N^{\rm jet}}{d^2 p_T \, d y} \> \bigg|_{y=0} &=&
  T_{AA} \> \frac{d \sigma^{\rm jet}}{d^2p_T \, dy} \> \bigg|_{y=0} 
  \nonumber \\[1.0em]
  &=& K \frac{C}{(1+p_T/B)^{\beta}} \> 
  \label{eq:para}
\end{eqnarray}
for the number of quarks, antiquarks and gluons at $y=0$. 
For SPS it is necessary to include a multiplicative term 
$(1-p_T/8.7 { \rm GeV})^\delta$ to the right hand side of Eq.~(\ref{eq:para}) 
in order to describe the $p_T$-spectrum close to the kinematic limit of 
8.7 GeV. Here 
$T_{AA}=9A^2/8\pi R_\perp^2$ is the nuclear thickness for zero impact 
parameter.
Values for the parameters $B$, $C$, $\beta$ and $\delta$ can be found in Table 
\ref{tab:minijets}. We note that the rapidity distributions of the
partons are quite flat near $y=0$ at RHIC and LHC energies. We have numerically
verified that assuming a flat rapidity spectrum at SPS energies introduces 
an error only of the order of 10--20\% as the expressions favour 
contributions from rapidities near $y=0$. 
In any case the yield at SPS energies is very low.

\begin{figure}[tb]
  \begin{center}  
  \epsfig{file=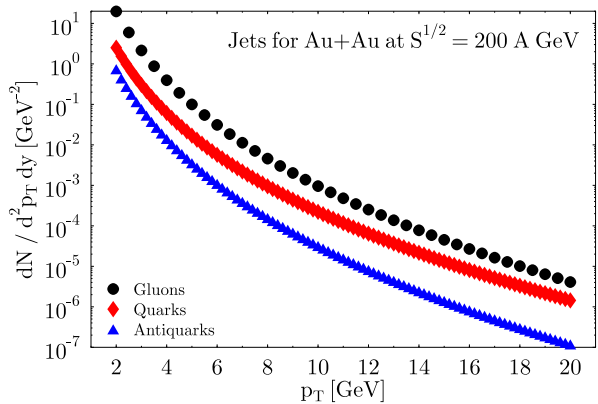,width=8.6cm}
  \epsfig{file=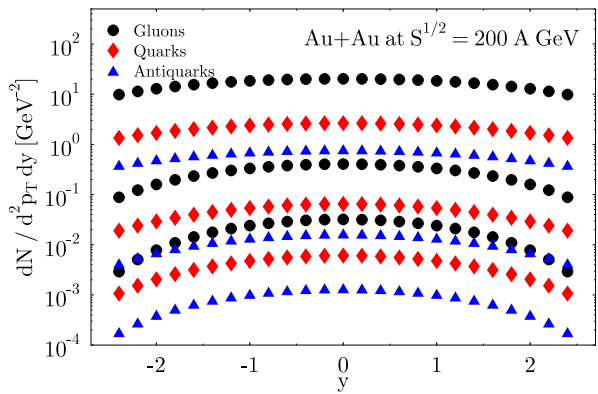,width=8.6cm}
  \caption{Same as Fig.~\ref{fig:jetsps} for central Au+Au at 
  $\sqrt{s_{NN}}=200$ GeV at RHIC.}
  \label{fig:jetrhic}
  \end{center}
\end{figure}

\begin{figure}[tb]
  \begin{center}
  \epsfig{file=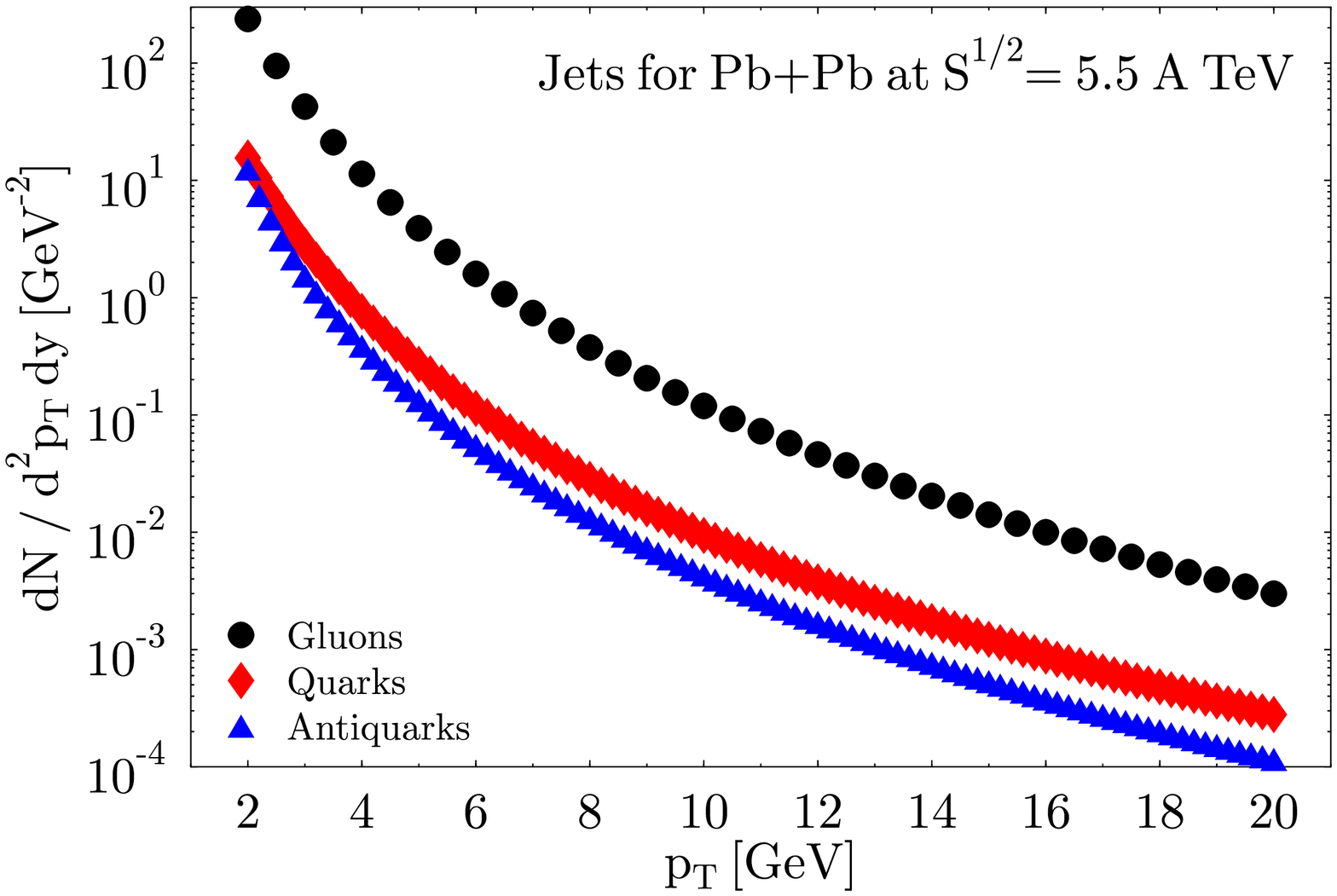,width=8.6cm} 
  \epsfig{file=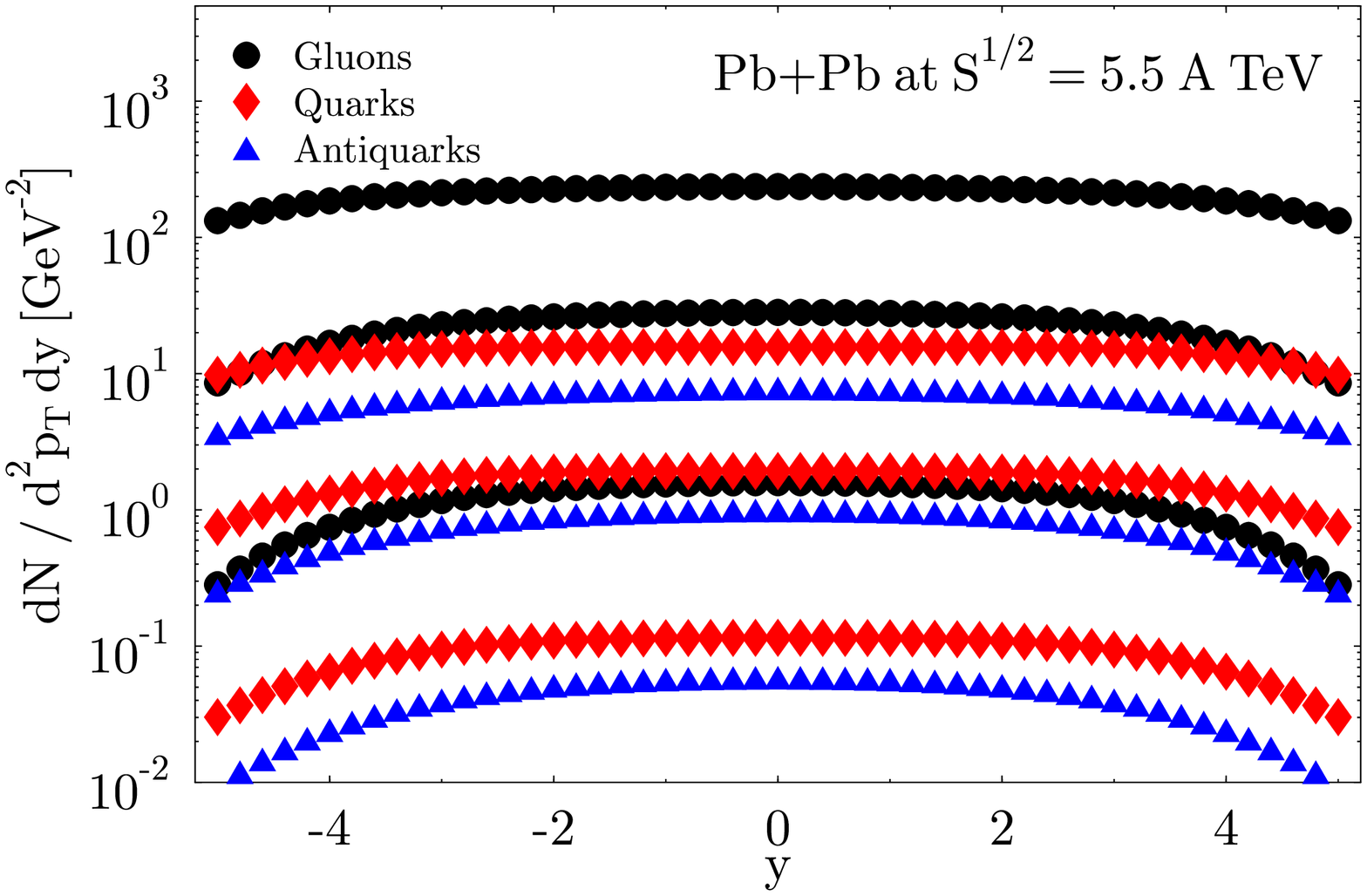,width=8.6cm} 
  \caption{Same as Fig.~\ref{fig:jetsps} for central Pb+Pb at 
  $\sqrt{s_{NN}}=5.5$ TeV at LHC.}
  \label{fig:jetlhc}
  \end{center}
\end{figure}

The  cross-section for the Drell-Yan process is given by,
\begin{multline}
  \frac{d\sigma}{dM^2\,dy} = \frac{4 \pi \alpha^2}{9M^4}\,\sum_{q} 
   e_q^2 \left[  x_1 f^A_q(x_1,M^2) x_2 f^B_{\overline{q}}(x_2,M^2) \right. \\
   \left. +(q \leftrightarrow \overline{q})\right]
\end{multline}
Here the sum runs again over the quark flavours, and
\begin{eqnarray}
x_1&= &M e^y/\sqrt{s_{NN}}, \nonumber\\
x_2 & = & M e^{-y}/\sqrt{s_{NN}}.
\end{eqnarray}
These are scaled by $T_{AA}$ similar to Eq.~(\ref{eq:para}) 
to obtain $dN/dM^2 \,dy$
for the Drell-Yan production of dileptons.

\section{Results and discussion}

\begin{figure}[tb]
  \begin{center}
  \epsfig{file=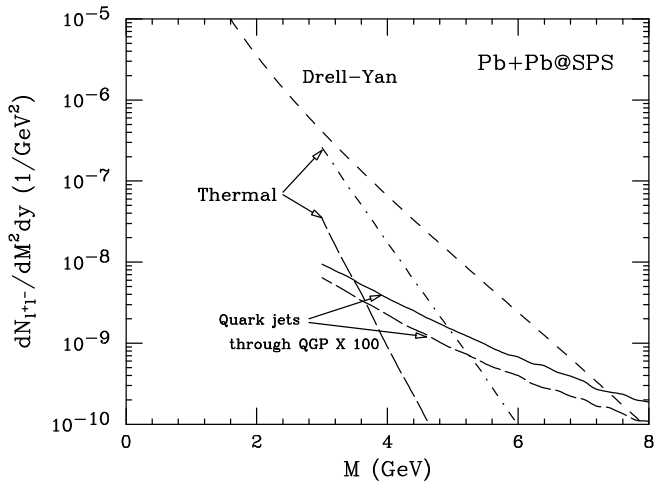,width=8.6cm} 
  \caption{Dilepton spectrum for Pb+Pb at $\sqrt{s_{\rm NN}}
  =17.4$ GeV at SPS.  The long dashed curves give the 
results when the formation time $\tau_0$ is raised to 0.50 fm/$c$,
thus lowering the temperature.}
  \label{fig:dilsps}
  \end{center}
\end{figure}

\begin{figure}[tb]
  \begin{center}  
  \epsfig{file=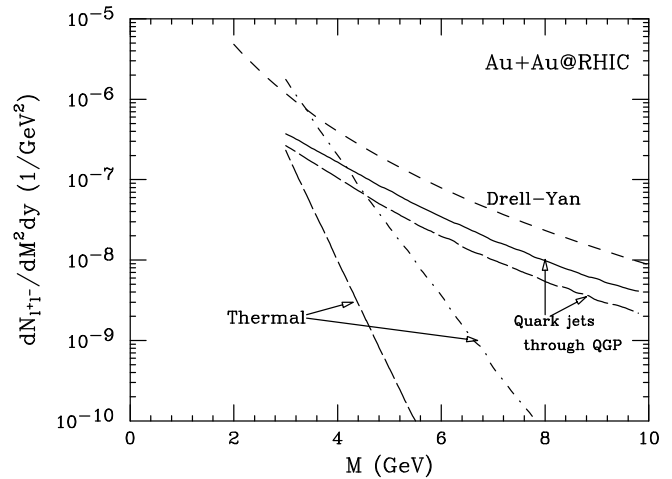,width=8.6cm} 
  \caption{Same as Fig.~\ref{fig:dilsps} for central Au+Au at 
  $\sqrt{s_{NN}}=200$ GeV at RHIC.}
  \label{fig:dilrhic}
  \end{center}
\end{figure}

\begin{figure}[tb]
  \begin{center}
  \epsfig{file=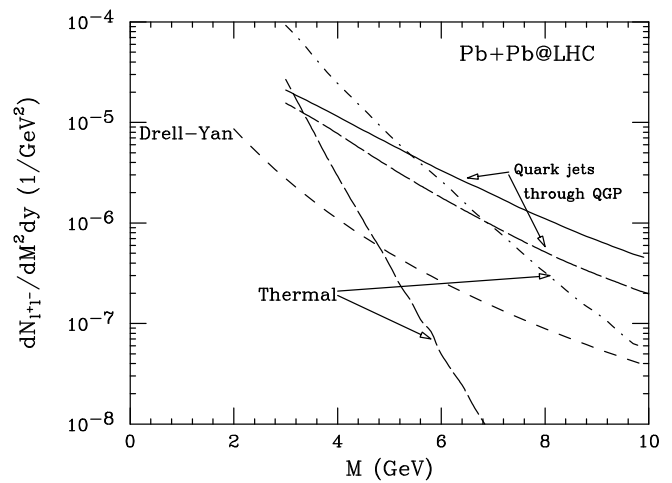,width=8.6cm} 
  \caption{Same as Fig.~\ref{fig:dilsps} for central Pb+Pb at 
  $\sqrt{s_{NN}}=5.5$ TeV at LHC.}
  \label{fig:dillhc}
  \end{center}
\end{figure}

In Figs.~\ref{fig:dilsps}, \ref{fig:dilrhic} and \ref{fig:dillhc} we plot the 
results for thermal dileptons, dileptons from the Drell-Yan process, 
and the dileptons from the passage of quark jets through the plasma 
for SPS, RHIC and LHC respectively. Note that gluon jets will contribute only
at higher order.

At SPS energies, we recover (Fig.~\ref{fig:dilsps})
 the well known result that the large mass 
dileptons have their origin predominantly in the Drell-Yan process. 
Increasing the formation time from 0.20 fm/$c$ to 
0.50 fm/$c$ --- and thus lowering the
initial temperature by 100 MeV --- drastically alters the thermal 
production (from the dash-dotted curve to the long-dashed one) while
the yield from the proposed jet-plasma interaction, even though
essentially negligible, is reduced by a factor of $\approx$ 2 (from the solid
line to the long-dashed one).

The jet-plasma interaction starts playing an interesting role at
RHIC energies (Fig.~\ref{fig:dilrhic}), as now the corresponding
yield is about only one third of the Drell-Yan contribution, and 
is much larger than the thermal contribution. Again
lowering the initial temperature (now by about 150 MeV) by increasing
the formation time to 0.50 fm/$c$ further enhances the importance 
of the yield due to jet-plasma interaction. We add that this production
is of the same order as that attributed to secondary-secondary 
quark-antiquark annihilation in a dilepton production calculation done using an earlier version of
the parton cascade model \cite{gk:93}.

The much larger initial temperatures likely to be attained at the
LHC and the much larger (mini)jet production lead 
to an excess of large mass dileptons from jet-plasma interactions
which can be an order of magnitude larger (at $M=10$ GeV) than that due
to the Drell-Yan process. Again, reducing the initial temperature
by raising the formation time to 0.50 fm/$c$ reduces the jet-plasma
yield by about a factor of 2 while the thermal yield is reduced
far more. We recall that at LHC energies several calculations
(see e.g.~\cite{kms:92,SriMusMul:97,ramona}) have reported a
thermal yield larger than
the Drell-Yan production.
We find that the jet-plasma interaction enhances the large mass 
dilepton production considerably.

\begin{figure}[htb]
  \begin{center}  
  \epsfig{file=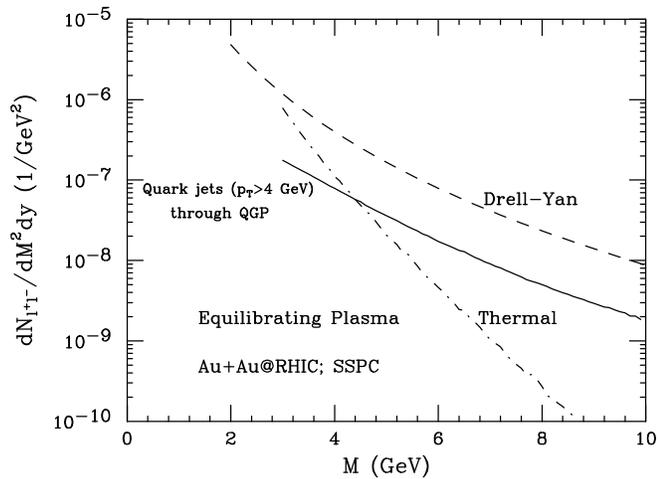,width=8.6cm} 
  \caption{Dilepton yield for central Au+Au at 
  $\sqrt{s_{NN}}=200$ GeV at RHIC assuming a chemically equilibrating plasma.}
  \label{fig:chemrhic}
  \end{center}
\end{figure}

The equilibrating plasma scenario discussed suggests that, while the 
quark and gluon fugacities may initially be less than unity, the 
initial temperature could be larger (see Table~\ref{tab:sspc}). This results
(see Fig.~\ref{fig:chemrhic})
in a situation where the jet-plasma yield remains large. It is larger than
the thermal yield, but still smaller than the Drell-Yan 
contribution at RHIC energies. 

\begin{figure}[htb]
  \begin{center}
  \epsfig{file=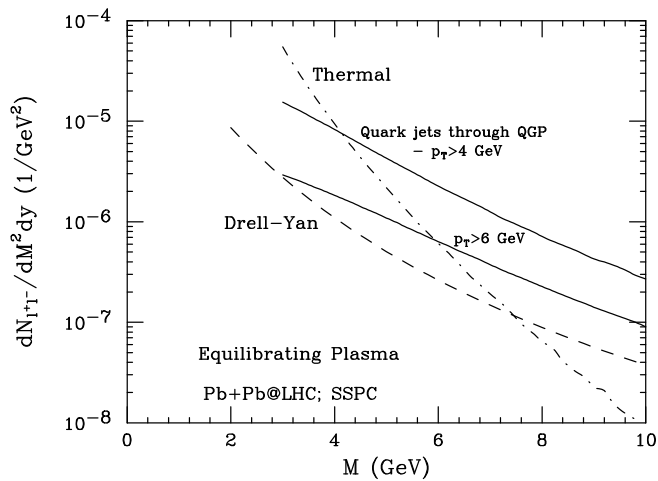,width=8.6cm} 
  \caption{Same as Fig.~\ref{fig:chemrhic} for central Pb+Pb at 
  $\sqrt{s_{NN}}=5.5$ TeV at LHC.}
  \label{fig:chemlhc}
  \end{center}
\end{figure}

The results at LHC energies (see Fig.~\ref{fig:chemlhc}) 
are particularly interesting because the
large mass yield is completely dominated by the jet-plasma interactions,
and this remains true even when only jets having $p_T > 6$ GeV are considered.


We see that the quark jets passing through the QGP give rise to a large
production of large mass dileptons, akin to the production of high 
energy photons through the same mechanism suggested earlier~\cite{fms}.
They should be absent in $pp$ collisions
and if no QGP is formed. 
We have noted that the results are
sensitive to the initial conditions and the time the jet spends in the
medium.

Before summarizing, we would like to discuss some interesting extensions of this work.
We have concentrated on quark jets. Gluon jets would give rise to dileptons 
through the process $qg \to q \gamma^*$. However, phase space 
considerations would tend 
to disfavour this process with respect to the channel considered in this work. 
But this contribution may still be large, 
especially as the gluonic jets are more numerous. This could also 
compensate for the additional $\alpha_s$ needed. However, a previous
 dynamical calculation did find
this contribution to be sub-leading \cite{gk:93}. 
The discussions in this work can be easily extended to diphotons~\cite{diphot}.
Connected to that is the interesting question of the transverse momentum 
spectrum of the dileptons, both from the Drell-Yan process and the 
jet-plasma interaction. No K-factors were  used  for the 
electromagnetic processes 
in this first baseline investigation. A detailed quantitative 
calculation will of course require
those, and we shall report on the results of those investigations in a 
future publication \cite{fgs:pt}.  We also add that one may in
principle have a energy and $p_T$ dependent $K$-factor~\cite{levai}
while estimating the jet-distributions. However, it is known that at higher
energies, where the discussed process becomes important, the $p_T$-dependence
of the $K$-factor is weak~\cite{levai}, when comparing NLO contributions
to the LO contributions. The precise value for the K-factor
is thus only a constant multiplicative factor and results for a different
value can be easily obtained.


A more serious modification of our results is expected from the fact that
jets passing through the QGP will loose energy. 
 We note though that quark jets, the focus of the present work,
lose less energy than the gluon jets. Furthermore, the number of 
quarks in the medium having a given transverse momentum $p_T$ drops 
rapidly as time passes, mainly due to the cooling (in the plasma) and also
due to the free streaming of the jets. This limits the production
of the dileptons mostly to the early times.
Several authors (see e.g., Ref.~\cite{sean})
 have discussed the effect of energy loss on the Drell-Yan production
of dileptons. This can be easily incorporated in the
treatment given here.

We have not compared our results to the correlated decay of charm
quarks, which provide a large contribution to dileptons at RHIC and 
LHC energies~\cite{ramona}, if the energy loss likely to be
suffered by the charm quarks due to collisions and radiations 
is ignored. If we assume that the charm quarks may lose energy
due to these processes~\cite{dipali}, then these contributions
will be suppressed~\cite{es}. 
This debate \cite{dima} has however not yet reached closure. A direct measurement could 
go a long way in settling this issue \cite{na6i}. 

We also recall that the source of dileptons discussed in the present
work corresponds to the so-called
secondary-secondary interactions in the parton cascade model,
discussed in Ref.~\cite{gk:93}. However, it was pointed out~\cite{eskola}
 that the procedure adopted in Ref.~\cite{gk:93} did not
 account for the fact that the
 final state partons emerging from a parton-parton collision continue to
participate in the development of the cascade by fragmenting and scattering
while  a virtual photon leaves the system. It will be worthwhile to
calculate the dilepton production within a (correctly implemented) 
parton cascade model~\cite{bms} with these considerations.

In brief, we have discussed a unique source of large mass dileptons
in relativistic collisions of heavy nuclei. It arises
from the passage of quark jets through the quark gluon plasma.
The contribution is found to be largest at LHC energies, moderate
at RHIC energies and negligible at SPS energies. Its detection
will provide a further proof of the existence of the QGP as well
as of multiple interactions suffered by the high energy partons
eventually leading to jet-quenching.

\acknowledgments  

This work was supported by DOE grants DE-FG02-96ER40945 and
DE-AC02-98CH10886 and the Natural Sciences and Engineering Research
Council of Canada.  RJF is supported by the Feodor Lynen program of 
the Alexander von Humboldt Foundation. We are indebted to Berndt
M\"uller for valuable comments.

\end{document}